\PassOptionsToPackage{colorlinks, citecolor=blue, urlcolor=black}{hyperref}    
\documentclass[traditabstract]{aa} 

\usepackage{natbib}
\bibpunct{(}{)}{;}{a}{}{,} 

\usepackage{graphicx}
\usepackage{subfigure}

\usepackage{savesym}
\usepackage{amsmath}
\savesymbol{iint}
\usepackage{txfonts}
\restoresymbol{TXF}{iint}

\usepackage{listings}
\usepackage{color}
                       
\usepackage{ifpdf}
\usepackage{url}
\ifpdf
\else
\fi

\definecolor{dkgreen}{rgb}{0,0.6,0}
\definecolor{gray}{rgb}{0.5,0.5,0.5}
\definecolor{mauve}{rgb}{0.58,0,0.82}

  \usepackage{courier}
 
\usepackage{caption}
\DeclareCaptionFont{white}{\color{white}}
\DeclareCaptionFont{black}{\color{black}}
\DeclareCaptionFont{bfblack}{\bf{\color{black}}}
\DeclareCaptionFormat{listing}{\hspace{-3pt}\colorbox[cmyk]{0.23, 0.15, 0.15,0.01}{\parbox{\textwidth}{\hspace{0pt}#1#2#3}}\vspace{-5mm}}
\newcommand{\nside}{N_{\rm side}}

\begin{document}

   \title{\texttt{S2LET}: A code to perform fast wavelet analysis on the sphere}


   \author{B.~Leistedt\inst{1}
          \and
          J.~D.~McEwen\inst{2}
          \and
          P.~Vandergheynst\inst{3}
          \and
          Y.~Wiaux\inst{4}
          }
   \institute{Department of Physics and Astronomy, University College
             London, London WC1E 6BT, UK \\
             \email{boris.leistedt.11@ucl.ac.uk}
         \and
             Department of Physics and Astronomy, University College
             London, London WC1E 6BT, UK \\
             Mullard Space Science Laboratory (MSSL), University College London, Surrey RH5 6NT, UK \\
             \email{jason.mcewen@ucl.ac.uk}
         \and
             Institute of Electrical Engineering, Ecole Polytechnique
             F{\'e}d{\'e}rale de Lausanne (EPFL), CH-1015 Lausanne,
             Switzerland \\
             \email{pierre.vandergheynst@epfl.ch}
         \and
             Institute of Electrical Engineering, Ecole Polytechnique
             F{\'e}d{\'e}rale de Lausanne (EPFL), CH-1015 Lausanne,
             Switzerland \\
             Department of Radiology and Medical Informatics, University of Geneva (UniGE), 
             CH-1211 Geneva, Switzerland\\
            Department of Medical Radiology, Lausanne University Hospital (CHUV), 
            CH-1011 Lausanne, Switzerland\\
            Institute of Sensors, Signals \& Systems, Heriot Watt University, Edinburgh EH14 4AS, UK\\
             \email{yves.wiaux@epfl.ch}
       }

   \date{Accepted August 2013}
 
   \abstract{We describe \texttt{S2LET}, a fast and robust implementation of the scale-discretised wavelet transform on the sphere.  Wavelets are constructed through a tiling of the harmonic line and can be used to probe spatially localised, scale-dependent features of signals on the sphere. 
    The reconstruction of a signal from its wavelets coefficients is made exact here through the use of a sampling theorem on the sphere.  Moreover, a multiresolution algorithm is presented to capture all information of each wavelet scale in the minimal number of samples on the sphere.  In addition \texttt{S2LET} supports the \texttt{HEALPix} pixelisation scheme, in which case the transform is not exact but nevertheless achieves good numerical accuracy.  The core routines of \texttt{S2LET} are written in \texttt{C} and have interfaces in \texttt{Matlab}, \texttt{IDL} and \texttt{Java}. Real signals can be written to and read from \texttt{FITS} files and plotted as Mollweide projections.  The \texttt{S2LET} code is made publicly available, is extensively documented, and ships with several examples in the four languages supported. At present the code is restricted to axisymmetric wavelets but will be extended to directional, steerable wavelets in a future release.}
   
   
   \maketitle
   
\section{Introduction}

Signals defined or measured on the sphere arise in numerous disciplines, where analysis techniques defined explicitly on the sphere are now in common use. In particular, wavelets on the sphere %
\citep{antoinevander1998, 
antoinevander1999, 
baldimarinucci2006needlets, 
Marinucci2008needlets, 
mcewen2006dirwavelets, 
narcowich:2006, 
starckmrs1, 
wiauxyvesvander2005c, 
wiaux2005correspondence,
mcewen2007dirwavelets,
mcewen2008dirwavelets,
yeo2008banks2sphere} have been applied very successfully to problems in astrophysics and cosmology, where data-sets are increasingly large and need to be analysed at high resolution in order to confront accurate theoretical predictions (e.g.\ %
\citealt{cobe2000gaussianityneedlets, 
delabrouille2012wmap7needletilc, 
cobe2001mexicannongaussianity, 
starckpires2012weaklensing, labatiestarck2012baos,
marinucci2008needletsbispectrum, 
mcewen2006bianchiwmap1, mcewenvielva2006iswdarkenergy, 
mcewen2007cosmoapplications, mcewenwiaux2007darkenergy,
marinucci2008cmbaniso, 
starckpires2006weaklensingwavelets, 
starck2010poissondenoising,  
vielva2004nongaussianitywmap1, vielva2006iswdarkenergy, 
vielvawiaux2006, vielvawiaux2007, wiauxvielva2006, wiauxvielva2008}).

While wavelet theory is well established in Euclidean space (see \emph{e.g.} \citealt{daubechies92}), multiple wavelet frameworks have been developed on the sphere, only a fraction of which lead to exact transforms in both the continuous and discrete settings. In fact, discrete methodologies \citep{schswe:sphere, swe:lift1, swe:lift2} achieve exactness in practice but may not lead to a stable basis on the sphere \citep{swe:lift2}. In the continuous setting several constructions are theoretically exact, and have been combined with sampling theorems on the sphere to enable exact reconstruction in the discrete setting also. In particular, scale-discretised wavelets \citep{mcewen2008dirwavelets} lean on a tiling of the harmonic line to yield an exact wavelet transform in both the continuous and discrete settings.  In the axisymmetric case, the scale-discretised wavelets reduce to needlets \citep{narcowich:2006,baldimarinucci2006needlets, Marinucci2008needlets}, which were developed independently using an analogous tiling of the harmonic line. {Similarly, the isotropic undecimated wavelet transform (UWT) developed by \citep{starckmrs1} exploits B-splines of order 3 to cover the harmonic line with filters with greater overlap but nevertheless compact support.}

In this paper we describe the new publicly available {\tt S2LET}\footnote{\url{http://www.s2let.org/}} code to perform the scale-discretised wavelet transform of complex signals on the sphere. At present {\tt S2LET} is restricted to axisymmetric wavelets (i.e. azimuthally symmetric when centred on the poles) and includes generating functions for axisymmetric scale-discretised wavelets \citep{mcewen2008dirwavelets}, needlets \citep{narcowich:2006,baldimarinucci2006needlets, Marinucci2008needlets} {and B-spline wavelets \citep{starckmrs1}}. We intend to extend the code to directional, steerable wavelets and spin functions in a future release. The core routines of {\tt S2LET} are written in {\tt C}, exploit fast algorithms on the sphere, and have interfaces in {\tt Matlab}, {\tt IDL} and {\tt Java}.

We note that many very useful public codes are already available to compute wavelet transforms on the sphere, including isotropic undecimated wavelet, ridgelet and curvelet transforms\footnote{\url{http://jstarck.free.fr/mrs.html}} \citep{starckmrs1}, invertible filter banks\footnote{\url{https://sites.google.com/site/yeoyeo02}} \citep{yeo2008banks2sphere}, needlets ({\tt NeedATool}\footnote{\url{http://www.fisica.uniroma2.it/~pietrobon/}}; \citealt{2010needatool}) and scale-discretised wavelets ({\tt S2DW}\footnote{\url{http://www.spinsht.org/}}; \citealt{mcewen2008dirwavelets}).  {{\tt S2LET} aims primarily to provide a fast and flexible implementation of the scale-discretised transform with exact reconstruction on the sphere using the sampling theorem of \cite{mcewen2011novelsampling}, although it has also been extended to support some of the features of these other codes. Furthermore, particular attention has been paid in the development of {\tt S2LET} to prove a user-friendly code, supporting multiple programming languages, and which is extensively documented. } 

The remainder of this article is organised as follows. In section 2 we detail the construction of scale-discretised axisymmetric wavelets and the corresponding exact scale-discretised wavelet transform on the sphere. In section 3 we describe the {\tt S2LET} code, including implementation details, computational complexity and numerical performance. We present a number of simple examples using {\tt S2LET} in section 4, along with the code to execute them. We conclude in section 5.

\section{Wavelets on the sphere}

{We review the construction of scale-discretised wavelets on the sphere through tiling of the harmonic line \citep{mcewen2008dirwavelets}.  Directional, steerable wavelets were also considered by \cite{mcewen2008dirwavelets}, however we restrict our attention to axisymmetric wavelets here.  Furthermore, the use of a sampling theorem on the sphere guarantees that spherical harmonic coefficients capture all the information content of band-limited signals, resulting in theoretically exact harmonic and wavelet transforms.  One may alternatively adopt samplings of the sphere for which exact quadrature rules do not exist, such as {\tt HEALPix} \citep{HEALPix}, but which nevertheless exhibit other useful properties, leading to numerically accurate but not theoretically exact transforms.}

\subsection{Harmonic analysis on the sphere}

The spherical harmonic decomposition of a square integrable signal $f \in L^2(S^2)$ on the two-dimensional sphere $S^2$ reads 
\begin{equation}
	f(\omega) = \sum_{\ell = 0}^\infty \sum_{m = -\ell}^\ell f_{\ell m} Y_{\ell m}(\omega), \label{harmonicssynthesis}
\end{equation}
where $Y_{\ell m}$ are the spherical harmonic functions, which form the canonical orthogonal basis on $S^2$.  The spherical harmonic coefficients $f_{\ell m}$, with $\ell\in \mathbb{N}$ and $m\in \mathbb{Z}$ such that $|m|\leq \ell$, form a dual representation of the signal $f$ in the harmonic basis on the sphere. The angular position $\omega=(\theta, \phi)\in S^2$ is specified by colatitude $\theta \in [0,\pi]$ and longitude $\phi \in [0,2\pi)$. The spherical harmonic coefficients are given by 
\begin{equation}
	f_{\ell m} = \langle f | Y_{\ell m} \rangle = \int_{S^2} {\rm d}\Omega(\omega) f(\omega) Y^*_{\ell m}(\omega), \label{harmonicsanalysis}
\end{equation}
with the surface element ${\rm d}\Omega(\omega) = \sin\theta {\rm d}\theta {\rm d}\phi$. We consider band-limited signals in the spherical harmonic basis, with band-limit $L$ if $f_{\ell m} = 0, \ \forall \ell \geq L$. For band-limited signals sampling theorems can be invoked so that both forward and inverse transforms can be reduced to finite summations that are theoretically exact.  Sampling theorems effectively encode a quadrature rule for the exact evaluation of integrals on the sphere from a finite set of sampling nodes.  Various sampling theorems exist in the literature (e.g.\ \citealt{driscollhealy1994, Healy96fftsfor, mcewen2011novelsampling}). In this work we adopt the \cite{mcewen2011novelsampling} sampling theorem (hereafter MW), which is based on an equiangular sampling scheme and, for a given band-limit $L$, requires the lowest number of samples on the sphere of all sampling theorems, namely $(L-1)(2L-1)+1\sim2L^2$ samples (for comparison $\sim4L^2$ samples are required by \cite{driscollhealy1994}).  Fast algorithms to compute the corresponding spherical harmonic transform scale as $\mathcal{O}(L^3)$ and are numerically stable to band-limits of at least $L = 4096$ \citep{mcewen2011novelsampling}. {The {{\tt GLESP}} pixelisation scheme \citep{glesp} also provides a sampling theorem based on the Gauss-Legendre quadrature, and could be used in place of the MW sampling theorem. However, {{\tt GLESP}} uses more samples than Gauss-Legendre quadrature requires, which may lead to an overhead when considering large band-limits and numerous wavelet scales. We focus on the MW sampling scheme to obtain a theoretically exact transform. Alternative sampling schemes that are not based on sampling theorems also exist such as {\tt HEALPix} \citep{HEALPix}, which is supported by  {\tt S2LET}, {\tt MRS} and {\tt Needatool}. {\tt HEALPix} does not lead to exact transforms on the sphere but the resulting approximate transforms nevertheless achieve good accuracy and benefit from other practical advantages, such as equal-area pixels.}

\subsection{Scale-discretised wavelets on the sphere}

{The scale-discretised wavelet transform allows one to probe spatially localised, scale-dependent content in the signal of interest \mbox{$f\in L^2(S^2)$}.} The $j$-th wavelet scale $W^{\Psi^{j}}\hspace{-1mm}\in L^2(S^2)$ is defined as the convolution of $f$ with the wavelet $\Psi^{j}\in L^2(S^2)$:
\begin{align}
  W^{\Psi^{j}}(\omega)  &\equiv (f \star \Psi^{j})(\omega)
  \equiv \langle f |  \mathcal{R}_\omega \Psi^{j} \rangle  \nonumber \\
  &\equiv \int_{S^2} {\rm d}\Omega(\omega^\prime) f(\omega^\prime) ( \mathcal{R}_\omega \Psi^{j} )^*(\omega^\prime),
\label{wav1}
\end{align} 
where ${}^*$ denotes complex conjugation. Convolution on the sphere is defined by the inner product of $f$ with the rotated wavelet $\mathcal{R}_\omega \Psi^{j}$.  We restrict our attention to axisymmetric wavelets, i.e.\ wavelets that are azimuthally symmetric when centred on the poles.  Consequently, the rotation operator $\mathcal{R}_\omega$ is parameterised by angular position $\omega=(\theta,\phi)$ only and not also orientation\footnote{As already noted, the extension to directional scale-discretised wavelets has been derived by \cite{mcewen2008dirwavelets}.  At present the {\tt S2LET} code supports axisymmetric wavelets only; directional wavelets will be added in a later release.}. %
For the axisymmetric case the spherical harmonic decomposition of $W^{\Psi^{j}}$ is then simply given by a weighted product in harmonic space:
\begin{equation}
	{W}^{\Psi^{j}}_{\ell m} =  \sqrt{ \frac{4\pi}{2\ell+1}} {f}_{\ell m} {\Psi}^{j*}_{\ell 0}, \label{wav2}
\end{equation}
where ${W}^{\Psi^{j}}_{\ell m} = \langle W^{\Psi^{j}} |Y_{\ell m} \rangle$, $f_{\ell m} = \langle f|Y_{\ell m} \rangle$ and ${\Psi}^{j}_{\ell 0}\delta_{m 0} = \langle \Psi^{j}|Y_{\ell m} \rangle$, and where $\delta_{m 0}$ is the Kronecker delta symbol.  

The wavelet coefficients extract the detail information of the signal only; a scaling function and corresponding scaling coefficients must be introduced to represent the low-frequency (\mbox{low-$\ell$}), approximate information of the signal.  The scaling coefficients $W^\Phi \in L^2(S^2)$ are defined by the convolution of $f$ with the scaling function $\Phi\in L^2(S^2)$:
\begin{equation}
W^\Phi(\omega)  \equiv (f \star \Phi)(\omega) = \langle f | \mathcal{R}_\omega \Phi \rangle, \label{wav3}
\end{equation}
or in harmonic space,
\begin{equation}
	{W}^\Phi_{\ell m} =  \sqrt{ \frac{4\pi}{2\ell+1}} {f}_{\ell m} {\Phi}^{*}_{\ell 0}, \label{wav4}
\end{equation}
where ${W}^{\Phi}_{\ell m} = \langle W^{\Phi} |Y_{\ell m} \rangle$ and ${\Phi}_{\ell 0}\delta_{m 0} = \langle \Phi|Y_{\ell m} \rangle$. 

Provided the wavelets and scaling function satisfy an admissibility property (defined below), the function $f$ may be reconstructed exactly from its wavelet and scaling coefficients by
\begin{align}
	\quad f(\omega) = & \int_{S^2} {\rm d}\Omega(\omega^\prime) W^{\Phi}(\omega^\prime)(\mathcal{R}_{\omega^\prime} \Phi)(\omega) \nonumber  \\
	&+ \ \sum_{j=J_0}^{J}  \int_{S^2} {\rm d}\Omega(\omega^\prime) W^{\Psi^{j}}(\omega^\prime)(\mathcal{R}_{\omega^\prime} \Psi^{j})(\omega), 
\end{align}
or equivalently in harmonic space by
\begin{equation}
	 {f}_{\ell m } = \sqrt{ \frac{4\pi}{2\ell+1}} {W}^{\Phi}_{\ell m} {\Phi}_{\ell 0} \  + \ \sqrt{ \frac{4\pi}{2\ell+1}} \sum_{j=J_0}^{J} {W}^{\Psi^{j}}_{\ell m} {\Psi}^{j}_{\ell 0}. \label{waveletsynthesis}
\end{equation}
The parameters $J_0$, $J$ define the lowest and highest scales $j$ of the wavelet decomposition and must be defined consistently to extract and reconstruct all the information content of $f$. These parameters depend on the construction of the wavelets and scaling function and are defined explicitly in the next paragraphs. The admissibility condition under which a band-limited function $f$ can be decomposed and reconstructed exactly is given by the following resolution of the identity:
\begin{equation}
	\frac{4\pi}{2\ell+1} \left( |{\Phi}_{\ell 0}|^2 + \sum_{j=J_0}^{J} |{\Psi}^{j}_{\ell 0}|^2 \right) \ = \ 1, \quad \forall \ell . \label{identity}
\end{equation}

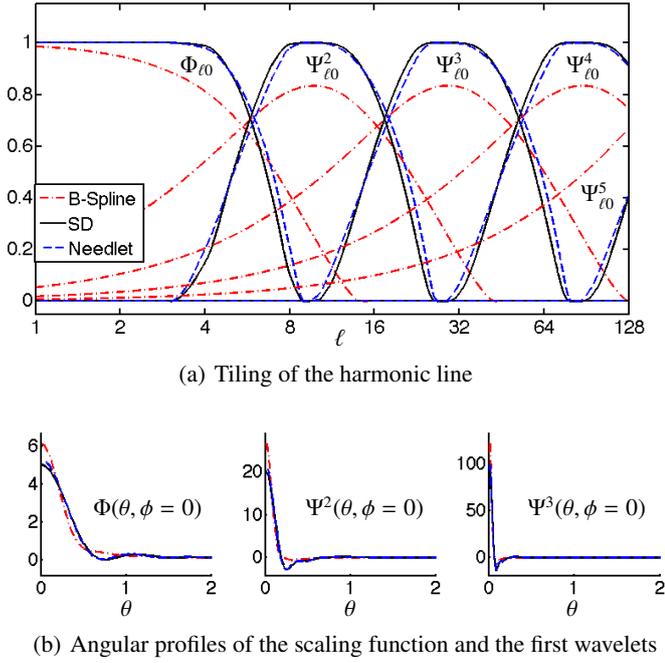
\begin{figure}
\centering
\setlength{\unitlength}{.5in}
\begin{picture}(7,4.4)(0,0)
\put(0.,0.5){\subfigure[Tiling of the harmonic line]{\includegraphics[trim = 2.9cm 0cm 2.3cm 0.5cm, clip, width=8.4cm]{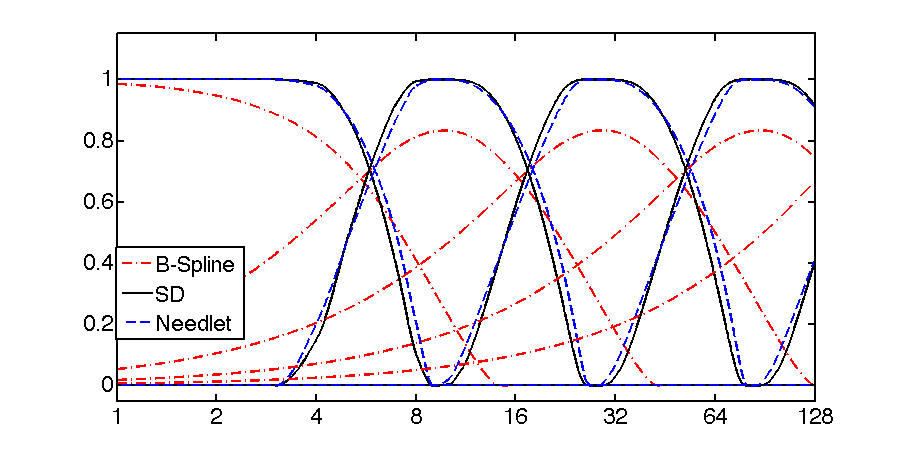}}}
\put(3.4,0.6){$\ell$}
\put(1.8,3.45){$\Phi_{\ell 0}$}
\put(3.1,3.45){$\Psi^2_{\ell 0}$}
\put(4.45,3.45){$\Psi^3_{\ell 0}$}
\put(5.8,3.45){$\Psi^4_{\ell 0}$}
\put(5.95,2.1){$\Psi^5_{\ell 0}$}
\end{picture}
\begin{picture}(7,2.7)(0,0)
\put(0.2,0.4){\subfigure[Angular profiles of the scaling function and the first wavelets]{\includegraphics[trim = 3.5cm -1.2cm 2.8cm 0.0cm, clip, width=8.4cm]{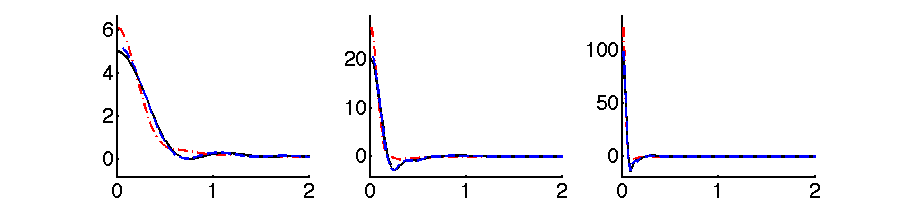}}}
\put(0.9,1.55){$\Phi(\theta,\phi=0)$}
\put(3.1,1.55){$\Psi^2(\theta,\phi=0)$}
\put(5.4,1.55){$\Psi^3(\theta,\phi=0)$}
\put(1.18,0.5){$\theta$}
\put(3.5,0.5){$\theta$}
\put(5.8,0.5){$\theta$}
\end{picture}
\caption{{Wavelets and scaling function constructed with the scale-discretised (SD), needlet and B-spline generating functions \citep{mcewen2008dirwavelets, Marinucci2008needlets, starckmrs1} with parameters $\lambda=3$ and $J_0=2$ and for band-limit $L=128$. The tiling is shown in the top panel, and the profiles of the reconstructed wavelets in the bottom panel. }}
\label{fig:kernels}
\end{figure}

\begin{figure}
\centering
\setlength{\unitlength}{.5in}
\begin{picture}(7,4.5)(0,0)
\put(0.2,0.4){\includegraphics[trim = 3.0cm 2.1cm 2.cm 1.0cm, clip, width=8.cm]{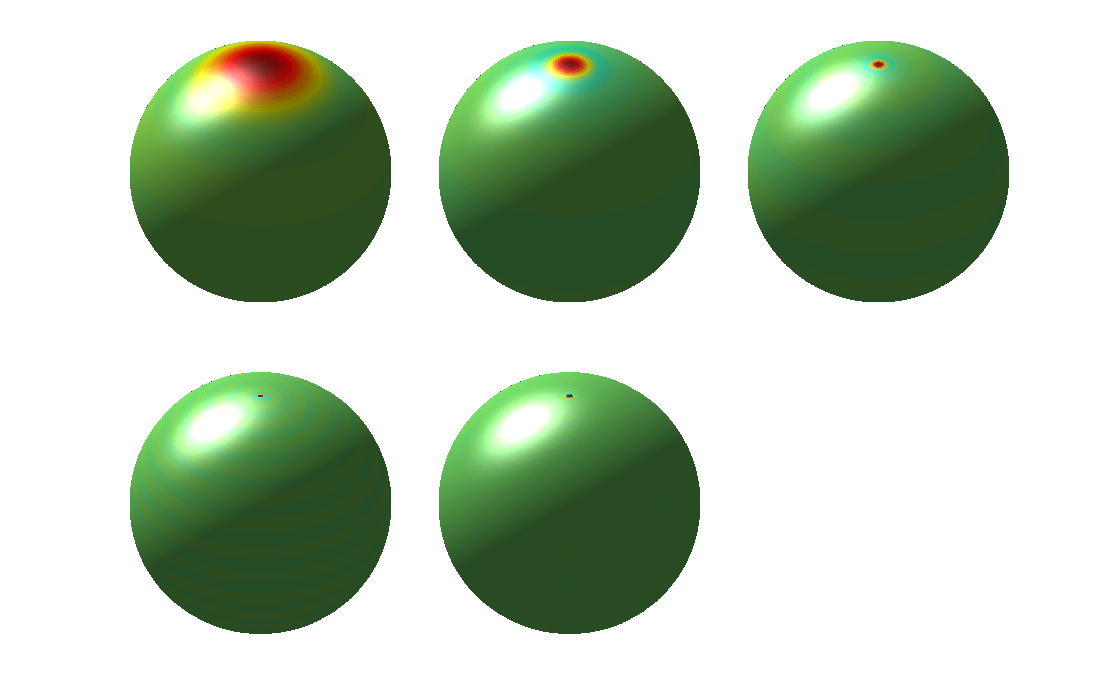}}
\put(1.15,2.35){$\Phi(\omega)$}
\put(3.15,2.35){$\Psi^2(\omega)$}
\put(5.2,2.35){$\Psi^3(\omega)$}
\put(1.15,0.16){$\Psi^4(\omega)$}
\put(3.2,0.16){$\Psi^5(\omega)$}
\end{picture}
\caption{Wavelets for scales $j \in \{2,3,4,5\}$ and scaling function constructed through a tiling of the harmonic line using scale-discretised functions, with parameters $\lambda=3$ and $J_0=2$ and for band-limit $L=128$. This plot was produced with a {\tt Matlab} demo included in {\tt S2LET}.}
\label{fig:kernelspix}
\end{figure}
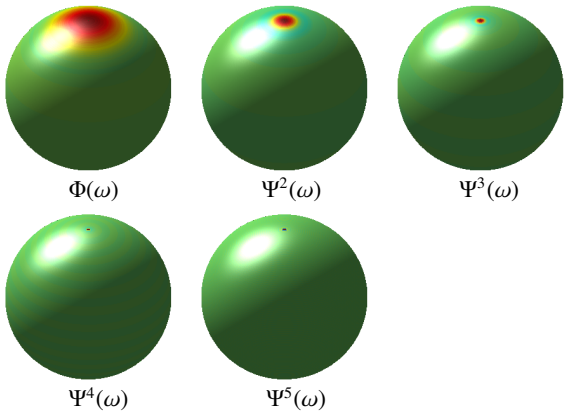

We are now in {a position} to define wavelets and a scaling function that satisfy the admissibility property.  In this paper, we use the smooth generating functions defined by \cite{mcewen2008dirwavelets} in order to tile the harmonic line. {Alternative definitions are also supported by {\tt S2LET}, as presented at the end of this section}. Consider the $C^{\infty}$ Schwartz function with compact support on $[-1,1]$:
\begin{equation}
	s(t) \equiv \left\{ \begin{array}{ll} \ e^{-\frac{1}{1-t^2}}, & t\in[-1,1] \\ \  0, & t \notin [-1,1]\end{array} \right. ,
\end{equation}
for $t \in \mathbb{R}$.  We introduce the positive real parameter $\lambda\in\mathbb{R}^+_*$ to map $s(t)$ to 
\begin{equation}
	s_\lambda(t) \equiv s\left( \frac{2\lambda}{\lambda-1} (t-1/\lambda)-1\right),
\end{equation}
which has compact support in $[1/\lambda, 1]$.  We then define the smoothly decreasing function $k_\lambda$ by
\begin{equation}
	 k_\lambda(t) \equiv \frac{\int_{t}^1\frac{{\rm d}t^\prime}{t^\prime}s_\lambda^2(t^\prime)}{\int_{1/\lambda}^1\frac{{\rm d}t^\prime}{t^\prime}s_\lambda^2(t^\prime)}, \label{smoothscaling}
\end{equation}
which is unity for $t<1/\lambda$, zero for $t>1$, and is smoothly decreasing from unity to zero for $t \in [1/\lambda,1]$. 
We finally define the wavelet generating function by
\begin{equation}
	 \kappa_\lambda(t) \equiv \sqrt{ k_\lambda(t/\lambda) - k_\lambda(t) }
\end{equation}
and the scaling function generating function by
\begin{equation}
	 \eta_{\lambda}(t) \equiv \sqrt{ k_\lambda(t)} .
\end{equation}
The wavelets and scaling function are constructed from their generating functions to satisfy the admissibility condition given by Eqn.~(\ref{identity}). A natural approach is to define ${\Psi}^{j}_{\ell m}$ from the generating functions $\kappa_\lambda$ to have support on $[\lambda^{j-1},\lambda^{j+1}]$, yielding
\begin{equation}
	{\Psi}^{j}_{\ell m} \equiv \sqrt{ \frac{2 \ell+1}{4\pi}}  \ \kappa_\lambda\left(\frac{\ell}{\lambda^j}\right) \delta_{m0}.
\end{equation}
For these wavelets Eqn.~(\ref{identity}) is satisfied for $\ell \geq \lambda^{J_0}$, where $J_0$ is the lowest wavelet scale used in the decomposition. The scaling function $\Phi$ is constructed to extract the modes that cannot be probed by the wavelets (i.e.\ modes with $\ell < \lambda^{J_0}$):
\begin{equation}
	{\Phi}_{\ell m}  \equiv  \sqrt{ \frac{2 \ell+1}{4\pi}}  \ \eta_{\lambda}\left(\frac{\ell}{\lambda^{J_0}}\right)\delta_{m0} .
\end{equation}
To satisfy exact reconstruction, $J$ is set to ensure the wavelets reach the band-limit of the signal of interest, yielding \mbox{$J = \lceil \log_\lambda(L-1) \rceil$}. The choice of the lowest wavelet scale $J_0$ is arbitrary, provided that $0 \leq J_0 < J$. The wavelets and scaling function may then be reconstructed on the sphere through an inverse spherical harmonic transform.  {The harmonic tiling and real space representation of these wavelets are shown in Figure~\ref{fig:kernels} and Figure~\ref{fig:kernelspix} respectively.}

 {In addition to the scale-discretised generating functions \citep{mcewen2008dirwavelets}, {\tt S2LET} also supports the needlet functions \citep{Marinucci2008needlets}\footnote{{In our implementation of needlets we use a scaling function to represent the approximate information in the signal, which is not always included (e.g., {\tt NeedAtool}; \citealt{2010needatool}).}}, which yield a similar tiling of the harmonic line, as shown in Figure~\ref{fig:kernels}. The B-spline filters used to construct the isotropic undecimated wavelet transform \citep{starckmrs1} are also supported, as also shown in Figure~\ref{fig:kernels}.\footnote{{For the B-spline-based construction to probe approximately the same scales as the scale-discretised and needlet ones, we defined the generating functions as
\begin{eqnarray}
 	k_\lambda(t) &=& \frac{3}{2} B_3(2 \frac{t\lambda^{J-1}}{L}) \\
	B_3(x) &=& \frac{1}{12}( |x-2|^3 - 4|x-1|^3 + 6|x|^3 - 4|x+1|^3 + |x+2|^3),
 \end{eqnarray}
 so that the $j$th filter has (compact) support $[0, L/\lambda^{J-j-2}]$ and peaks at the same scales as the $j$-th scale-discretised and needlet filters obtained with the same parameters.}
} With these three constructions, the wavelets and scaling functions are well-localised both spatially on the sphere and also in harmonic space.  Consequently, the associated wavelet transforms on the sphere can be used to extract spatially localised, scale-dependent features in signals of interest.}

\section{The {\tt S2LET} code}

In this section we describe the {\tt S2LET} code.  We first introduce a multiresolution algorithm to capture each wavelet scale in the minimum number of samples on the sphere, which follows by taking advantage of the reduced band-limit of the wavelets for scales $j<J-1$.  This multiresolution algorithm reduces the computation cost of the transform considerably. We then provide details of the implementation, the computational complexity and the numerical accuracy of the scale-discretised wavelet transform {supported} in {\tt S2LET}. We finally outline {planned} future extensions of the code.

\subsection{Multiresolution algorithm}

\begin{figure}[h!]
\centering
\subfigure[Full-resolution scale-discretised wavelet transform]{\includegraphics[trim = 3.1cm 2.4cm 0.9cm 1.4cm, clip, width=8.5cm]{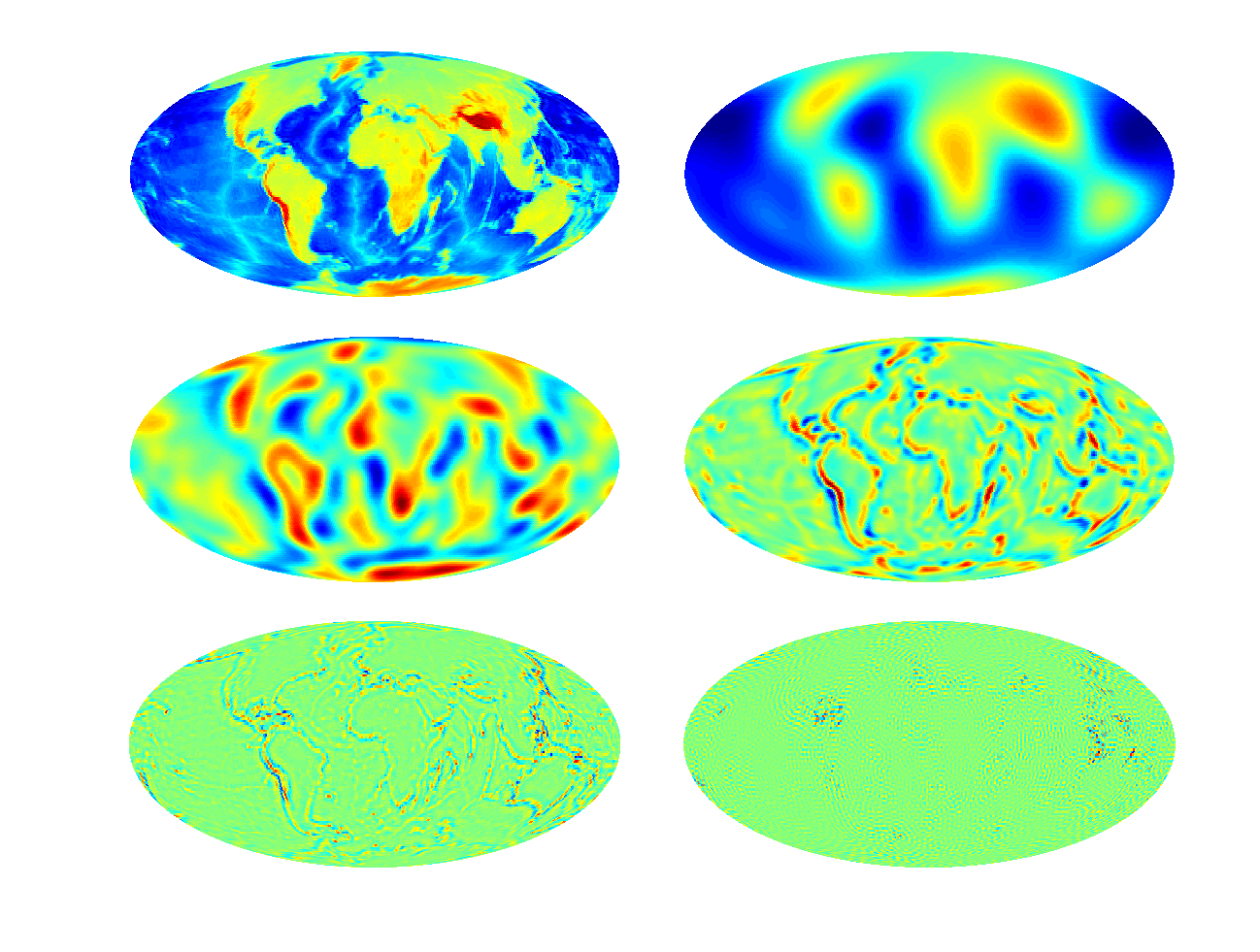}}
\subfigure[Multiresolution scale-discretised wavelet transform]{\includegraphics[trim = 3.1cm 2.4cm 0.9cm 1.4cm, clip, width=8.5cm]{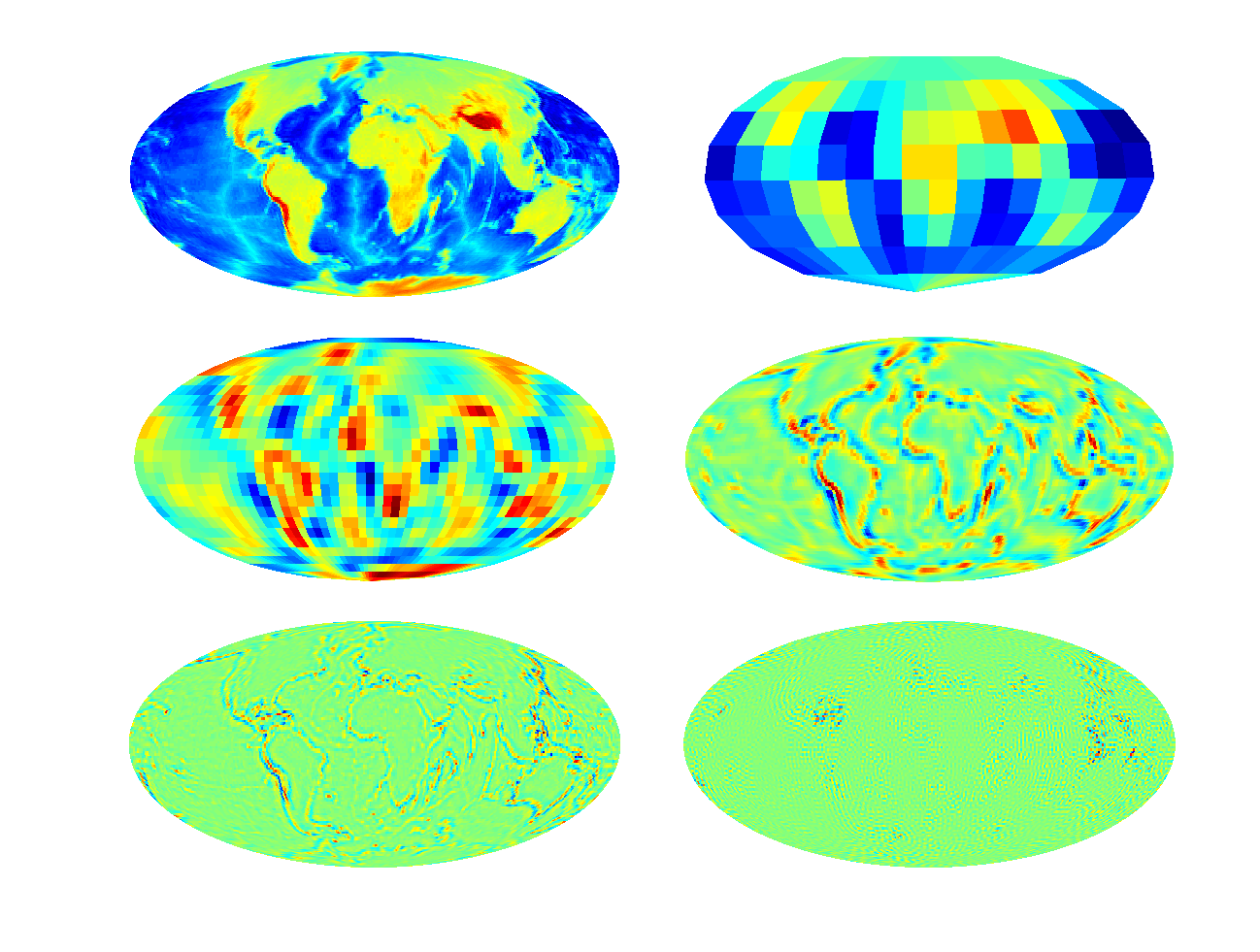}}
\caption{Scale-discretised wavelet transform of a band-limited topography map of the Earth for $\lambda=3$, $ J_0=2$ and $L=128$, i.e.\ with the scale-discretised wavelets shown in Figure~\ref{fig:kernelspix}. The wavelet transform decomposes the band-limited signal into wavelet coefficients that extract spatially localised, scale-dependent features. Since the wavelets for different scales $j$ have different band-limits, the wavelet coefficients can be reconstructed at lower resolution on the sphere for lower scales $j$. Panel~(a) shows the full-resolution wavelet transform of the topography map.  The original Earth topography map is shown in the top-left plot, the scaling coefficients are shown in the top-right plot, while the wavelet coefficients at scales $j\in\{2,3,4,5\}$ are shown left-to-right, top-to-bottom respectively in the remaining plots.  Panel~(b) shows the same decomposition but using the multiresolution algorithm.  The signals shown in panel~(b) contain the same information as in panel~(a) but represented in the minimal number of samples on the sphere.  These plots were produced by one of the many {\tt Matlab} demos provided with {\tt S2LET}.}
\label{fig:earth} 
\end{figure}

{In harmonic space, the wavelet coefficients are simply given by the weighted product of the spherical harmonic coefficients of $f$ and the wavelets, as expressed in Eqn.~(\ref{wav2}). Although the wavelet coefficients can be analysed at the same resolution as the signal $f$ (i.e., at full resolution), by construction they have different band-limits for different scales $j$, as shown in Figure~\ref{fig:kernels}. The reconstruction can thus be performed at lower resolution, without any loss of information if a sampling theorem is used. This approach yields a multiresolution algorithm where the wavelet coefficients are reconstructed with the minimal number of samples on the sphere: {the $j$-th wavelet coefficients have band-limit $k = \lambda^{j+1}$ when using the scale-discretised and needlet kernels, and $k = L / \lambda^{J-j-2}$ when using the B-splines}. When the MW sampling theorem is used, the wavelets are recovered on $(k-1)(2k-1)+1$ samples on the sphere. This approach leads to significant improvements in terms of speed and memory use compared to the full-resolution case, as shown in the next section. Figure~\ref{fig:earth} illustrates the use of the full-resolution and multiresolution transforms on a map of Earth topography data with the scale-discretised filters and the MW scheme. When adopting the {\tt HEALPix} sampling of the sphere, multiresolution can also be used. However {\tt HEALPix} does not rely on a sampling theorem and therefore the resolution for the reconstruction of each wavelet scale must be chosen heuristically and adapted to the desired accuracy. For example, in the {\tt MRS} code \citep{starckmrs1} it is chosen such that $\nside^j = k/2$. More detail on the accuracy of the wavelet transform with {\tt HEALPix} are provided below.}

\subsection{Implementation}

The core numerical routines of {\bf {\tt S2LET}} are implemented in {\tt C}.  By adopting a low level programming language such as {\tt C} for the implementation of the core algorithms, computational efficiency is optimised.  The {\tt C} library includes the full-resolution and \mbox{multi\-resolution} wavelet transforms, with specific optimisations for real signals in order to take advantage of all symmetries of the spherical harmonic transform. The wavelet transform is computed in harmonic space through Eqn.~(\ref{wav2}) and Eqn.~(\ref{wav4}), for the input parameters $(L, \lambda, J_0)$. To reconstruct signals on the sphere, by default {\tt S2LET} uses the exact spherical harmonic transform of the MW sampling theorem \citep{mcewen2011novelsampling} implemented in the {\tt SSHT}\footnote{\url{http://www.spinsht.org/}} code. In this case all transforms are theoretically exact and one can analyse and synthesise real and complex signals at floating-point precision. {{\tt S2LET} has been extended to also support the {\tt HEALPix} sampling scheme, in which case the transform is not theoretically exact but nevertheless achieves good numerical accuracy.}

We provide interfaces for the {\tt C} library in three languages: {\tt Matlab}, {\tt IDL} and {\tt Java}. The {\tt Matlab} and {\tt IDL} codes also include routines to read/write signals on the sphere stored in either {\tt HEALPix} {\tt FITS}\footnote{\url{http://fits.gsfc.nasa.gov/}} files or the {\tt FITS} file format used to stored MW sampled signals.  In addition, functionality to plot the Mollweide projection of real signals for both MW or {\tt HEALPix} samplings is included. The {\tt Java} interface includes an object-oriented representation of sampled maps, spherical harmonics and wavelet transforms. All routines and interfaces are well documented and illustrated with several examples for both the MW and {\tt HEALPix} samplings. These examples cover multiple combinations of parameters and types of signals. {\tt S2LET} requires {\tt SSHT}, which implements fast and exact algorithms to perform the forward and inverse spherical harmonic transforms corresponding to the MW sampling theorem \citep{mcewen2011novelsampling}. {\tt SSHT} in turn requires the {\tt FFTW}\footnote{\url{http://www.fftw.org/}} package for the computation of fast Fourier transforms. The fast spherical harmonic transforms implemented in {\tt SSHT} compute Wigner functions, and thus the spherical harmonic functions, through efficient recursion using either the method of \cite{trapani2006} or \cite{risbo1996}.  Here we present results using the recursion of \cite{risbo1996}. The fast spherical harmonic transform algorithms implemented in {\tt SSHT} scale as $\mathcal{O}(L^3)$ \citep{mcewen2011novelsampling}.

{Although primarily intended to perform the scale-discretised wavelet transform of \cite{mcewen2008dirwavelets}, {\tt S2LET} also supports the needlet and spline-based wavelet transforms developed by \cite{Marinucci2008needlets} and \cite{starckmrs1}. As shown in Figure~\ref{fig:kernels}, these generating functions yield the same number of wavelet scales (for the parameter choices described previously). However, with the scale-discretised and needlet generating functions the $j$-th wavelet scale has compact support in $[\lambda^{j-1},\lambda^{j+1}]$, whereas the support is much wider with the B-splines, i.e. $[0, L / \lambda^{J-j-2}]$ in the {\tt S2LET} implementation. As a consequence, when using the multiresolution algorithm the wavelet coefficients must be captured on a greater number of pixels than with the scale-discretised or needlet kernels, while probing approximately the same scales, as shown in Figure~\ref{fig:kernels}.} 

The complexity of the axisymmetric wavelet transform is dominated by spherical harmonic transforms since the wavelet transforms are computed efficiently in harmonic space, through Eqn.~(\ref{wav2}) and Eqn.~(\ref{wav4}) for the forward transform and through Eqn.~(\ref{waveletsynthesis}) for the inverse transform.  Given a band-limit $L$ and wavelet parameters $(\lambda, J_0)$, recall that the maximum scale is given by $J = \lceil \log_\lambda(L-1) \rceil$ and hence the wavelet transform (forward or inverse) involves \mbox{$(J-J_0+3)$} spherical harmonic transforms (one for the original signal, one for the scaling coefficients and \mbox{$(J-J_0+1)$} for the wavelet coefficients). If the scaling coefficients and all wavelet coefficients are reconstructed at full-resolution in real space, the axisymmetric wavelet transform scales as $\mathcal{O}((J-J_0+3){L^3})$. However, in the previous section we established a multiresolution algorithm that takes advantage of the reduced band-limit of the wavelets for scales $j<J-1$.  With the multiresolution algorithm with a sampling theorem, only the finest wavelet scales $j\in\{J-1,J\}$ are computed at maximal resolution corresponding to the band-limit of the signal. The complexity of the overall multiresolution wavelet transform is then dominated by these operations and effectively scales as $\mathcal{O}({L^3})$.

\subsection{Numerical validation}

\begin{figure}
\centering
\subfigure[Numerical accuracy of the wavelet transform]{\includegraphics[trim = 7.0cm 10.6cm 1cm 1cm, clip, width=9cm]{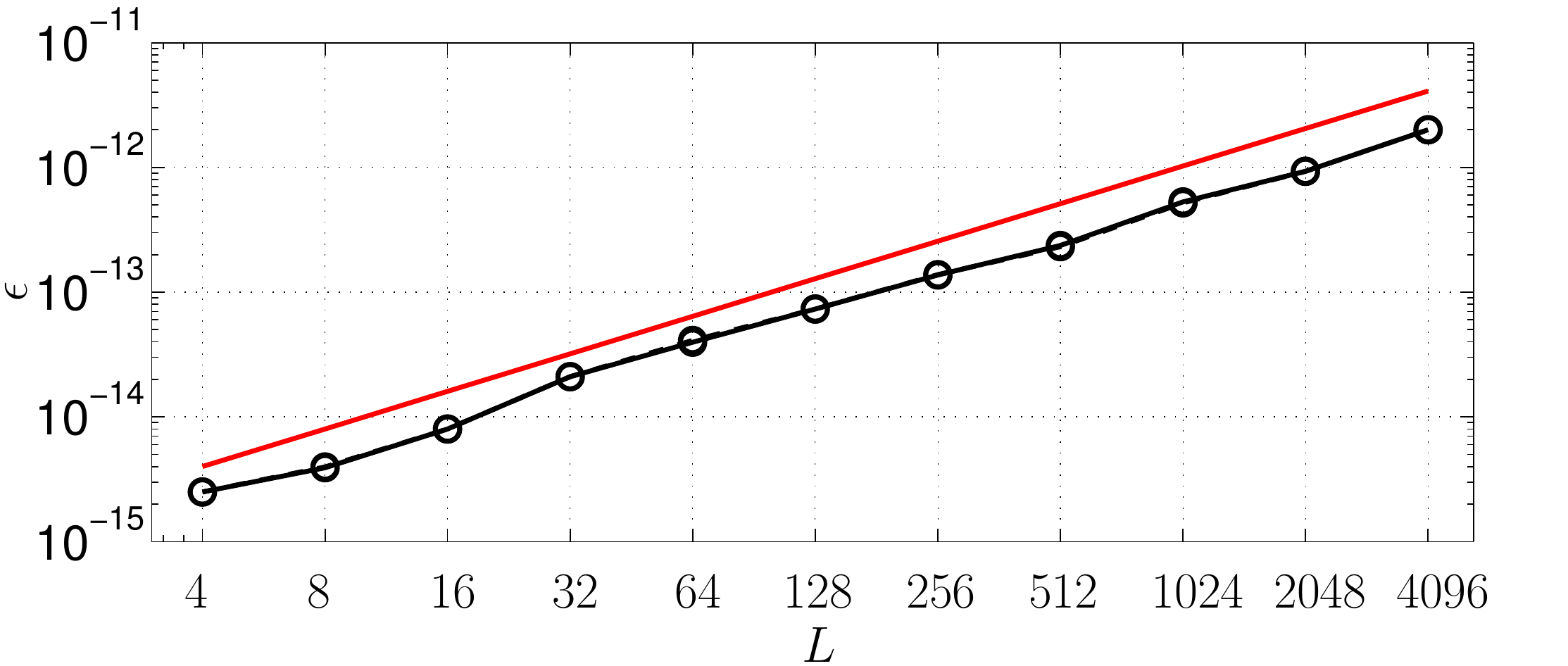}}
\subfigure[Computation time of the wavelet transform]{\includegraphics[trim = 7.0cm 0.5cm 1cm 11.5cm, clip, width=9cm]{pics/perfs.pdf}}
\caption{Numerical accuracy and computation time of the scale-discretised wavelet transform computed with {\tt {\tt S2LET}}. We consider $L=2^i$ with $i \in \{2,\ldots,10\}$, with parameters $\lambda = 2$, $J_0 =0$. These results are averaged over many realisations  of random band-limited signals and were found to be very stable. The scale-discretised transform is either performed at full-resolution (solid lines) or with the multiresolution algorithm (dashed lines). Very good numerical accuracy is achieved by both the full-resolution and multiresolution algorithms (which achieve indistinguishable accuracy), with numerical errors comparable to {floating-point} precision, found empirically to scale as $\mathcal{O}(L)$ as shown by the red line in panel~(a). Computation time scales as $\mathcal{O}(L^3)$ for both algorithms as shown by the red line in panel~(b), in agreement with theory.  The multiresolution algorithm is four to five times faster than the full-resolution approach for the band-limits considered.}
\label{fig:perfs}
\end{figure}

We {first} evaluate the performance of {\tt S2LET} in terms of accuracy and complexity using the MW sampling theorem, for which all transforms are theoretically exact. We show that {\tt S2LET} achieves floating-point precision and scales as detailed in the previous section. 

We consider band-limits \mbox{$L=2^i$} with \mbox{$i \in \{2,\ldots,10\}$} and generate sets of spherical harmonic coefficients $f_{\ell m}$ following independent Gaussian distributions $\mathcal{N}(0,1)$. We then perform the wavelet decomposition and reconstruct the harmonic coefficients, denoted by $f_{\ell m}^\textrm{rec}$.  We evaluate the accuracy of the transform using the error metric \mbox{$\epsilon = \max | f_{\ell m} - f_{\ell m}^\textrm{rec}|$}, which is theoretically zero since all signals are band-limited by construction. The complexity is quantified by observing how the computation time \mbox{$t_{\rm c} = [t_{\textrm{synthesis}} +  t_{\textrm{analysis}} ]/2$} scales with band-limit, where the synthesis and analysis computation times are specified by $t_{\textrm{synthesis}}$ and $t_{\textrm{analysis}}$ respectively.  Since we evaluate the wavelet transform in real space, a preliminary step is required to reconstruct the signal $f$ from the randomly generated $f_{\ell m}$. This step is not included in the computation time since its only purpose is to generate a valid band-limited test signal on the sphere. The analysis then denotes the decomposition of $f$ into wavelet coefficients $W^{{\Psi}^{j}}$ and scaling coefficients $W^\Phi$ on the sphere. The synthesis refers to recovering the signal $f^{\textrm{rec}}$ from these coefficients. The final step, which is not included in the computation time either, is to decompose $f^{\textrm{rec}}$ into harmonic coefficients $f^{\textrm{rec}}_{\ell m}$ in order to compare them with $f_{\ell m}$.  The stability of both $\epsilon$ and $t_{\rm c}$ is checked by averaging over hundreds of realisations of $f_{\ell m}$ for \mbox{$L=2^i$} with \mbox{$i \in \{2,\ldots,8\}$} and a few realisations with \mbox{$i \in \{9,10\}$}. The results proved to be very stable, i.e. the variances of the error and timing metrics are lower than 5\%. Recall that for given band-limit $L$ the number of samples on the sphere required by the exact quadrature is $(2L-1)(L-1)+1$. All tests were run on an Intel 2.0GHz Core i7 processor with 8GB of RAM. {On this machine, precision of floating point numbers is of the order of $\sim 10^{-16}$, and errors are expected to add up and accumulate when considering linear operations such as the spherical harmonic and wavelet transforms.} The accuracy and timing performance of the scale-discretised wavelet transform implemented in {\tt S2LET} are presented in Figure~\ref{fig:perfs}. {\tt S2LET} achieves very good numerical accuracy, with numerical errors comparable to {accumulated floating-point errors only}\footnote{{The {{\tt GLESP}} sampling adopted in {\tt MRS} also achieves floating point accuracy, although using many more pixels to capture the wavelet scales due to the greater band-limits of the spline-based kernels and the oversampling of the {\tt GLESP} scheme.}}. Moreover, the full-resolution and multiresolution algorithms are indistinguishable in terms of accuracy. However, the latter is four to five times faster than the former for the band-limits considered since only the wavelet coefficients for \mbox{$j\in\{J-1,J\}$} are computed at full-resolution.  As shown in Figure~\ref{fig:perfs}, computation time scales as $\mathcal{O}({L^3})$ for both algorithms, in agreement with theory.

{{\tt S2LET} can also be used with {\tt HEALPix}, in which case the accuracy of the spherical harmonic transform is critical to the accuracy of the wavelet transform (since {\tt HEALpix} does not rely on a sampling theorem it does not exhibit theoretically exact harmonic transforms, unlike {\tt SSHT} or {\tt GLESP}). The performances of the spherical harmonic transforms in {\tt HEALPix} and {{\tt GLESP}} have been widely studied in the past (see, e.g., \citealt{Libpsht2011, Glesp2011, Reinecke2013libsharp}), and that of the MW sampling were presented in \cite{mcewen2011novelsampling}. We do not compile the entirety of these results here, but we have reproduced the essential results on our machine; Table~\ref{tab:hpxacc} summarises the orders of accuracy of the {\tt HEALPix} iterative spherical harmonic transform. Using the same setup as previously, we calculated the maximum error on the spherical harmonic coefficients when performing the transform back and forth, averaged over the values of $\nside$, since the results were found to be sensitive only to the ratio $L / \nside$. Even with several iterations, which multiplies the number of transforms and thus computation time, the spherical harmonic transform in {\tt HEALPix} remains at least an order of magnitude less accurate than the MW and {\tt GLESP} counterparts (which, being both theoretically exact, achieve comparable performances, see \citealt{Libpsht2011, Reinecke2013libsharp, Glesp2011, mcewen2011novelsampling}). Since the wavelet transforms implemented in {\tt MRS} and {\tt Needatool} are also computed in harmonic space, their complexity and accuracy are dominated by that of the underlying spherical harmonic transforms. As a consequence, when adopting the {\tt HEALPix} scheme, {\tt S2LET}, {\tt MRS} and {\tt Needatool} achieve similar performances, resulting from the computation time and accumulated errors of $(J-J_0+1)$ {\tt HEALPix} spherical harmonic transforms. In the multiresolution case, the results depend on the resolution chosen to reconstruct each wavelet scale.
}

\begin{table}
	\centering
	\begin{tabular}{c|cccc}
		\hspace*{-4pt}$\max | f_{\ell m} - f_{\ell m}^\textrm{rec}|$\hspace*{-3pt}	& $L = \nside / 2$ & $L = \nside$ & \hspace*{-2pt}$L = 2\nside$ & \hspace*{-3pt}$L = 3\nside$ \\ 	\hline
		0 iteration & $\sim 10^{-6}$ & $\sim 10^{-4}$ & $\sim 10^{-2}$ & $\sim 10^{-1}$\\
		1 iterations & $\sim 10^{-10}$ & $\sim 10^{-7}$ & $\sim 10^{-3}$ & $\sim 10^{-1}$\\
		2 iterations & $\sim 10^{-14}$ & $\sim 10^{-10}$ & $\sim 10^{-5}$ & $\sim 10^{-1}$\\
		3 iterations & $\sim 10^{-14}$ & $\sim 10^{-13}$ & $\sim 10^{-6}$ & $\sim 10^{-1}$\\
		4 iterations & $\sim 10^{-14}$ & $\sim 10^{-14}$ & $\sim 10^{-7}$ & $\sim 10^{-1}$ \\ \hline
	\end{tabular}
	\caption{{Order of magnitude of the accuracy of the {\tt HEALPix} spherical harmonic transform, averaged over the parameter $\nside$.}}	
	\label{tab:hpxacc}
\end{table}

\subsection{Future extensions}

 In future work we plan to extend {\tt S2LET} to support directional, steerable wavelets on the sphere \citep{mcewen2008dirwavelets}. We also plan to exploit recent ideas leading to fast (spin) spherical harmonic transforms \citep{mcewen2011novelsampling} to yield faster algorithms than those developed by \cite{mcewen2008dirwavelets} and \cite{mcewen:2013:waveletsxv} to compute directional wavelet transforms on the sphere.  Finally, we intend to add support to analyse spin signals on the sphere {\citep[c.f.][]{Geller2008spinCMB, Geller2010spin1, Starck2009spinwavelets}}. In a future release, the code will also be parallelised, which will lead to further speed improvements. 
  The {\tt S2LET} code will thus be under active development with future releases forthcoming.  In any case, we hope this first version of the {\tt S2LET} code will prove useful for axisymmetric scale-discretised wavelet analysis on the sphere.  Indeed, the code has already been used as an integral part of the new exact flaglet wavelet transform on the ball \citep{flaglets}, the spherical space constructed by augmenting the sphere with the radial line.

\section{Examples}

The {\tt S2LET} code is extensively documented and ships with several examples in the four languages supported. In this section we present a subset of short examples, along with the code to execute them in order to demonstrate the ease of using {\tt S2LET} to perform wavelet transforms\footnote{Note that the code uses a slightly different notation compared to the equations of this article: $B$ refers to the wavelet scaling parameter (denoted $\lambda$ herein) and $J_{\rm min}$ to the first scale of the transform (denoted $J_0$ herein).}. {All examples were run with the scale-discretised wavelet generating functions.}

\subsection{Wavelet transform from the command line}

{\tt S2LET} includes ready-to-use high-level programs to directly decompose a real signal into wavelet coefficients. The inputs are a {\tt FITS} file containing the signal of interest and the parameters for the transform. The program writes the output coefficients in {\tt FITS} files in the same directory as the input file and with a consistent naming scheme. These commands are available for both {\tt HEALPix} and MW sampling schemes. For the MW sampling case illustrated in Example~\ref{codeBash1}, the wavelet transform is {theoretically} exact and the band limit corresponds to the resolution of the input map, which will be read automatically from the file. The transform may be performed in full-resolution or multiresolution by adjusting the multiresolution flag specified by the last parameter (respectively $0$ and $1$), and the output wavelet coefficients are computed at full and minimal resolution accordingly. For the case of a {\tt HEALPix} map, as illustrated in Example~\ref{codeBash2}, the band-limit must be supplied as the last parameter in the command. The output scaling and wavelet coefficients of a {\tt HEALPix} map are reconstructed and stored in {\tt FITS} files at the same resolution as the input map. For both MW and {\tt HEALPix} samplings the output coefficients may be read and plotted using the {\tt Matlab} or {\tt IDL} routines.

\begin{figure}[h!]\begin{minipage}{0.95\linewidth}\begin{lstlisting}[label=codeBash1, caption=Performing the forward (analysis) and inverse (synthesis) wavelet transform of a real signal (MW sampling) from the command line.]
>> ./bin/s2let_axisym_mw_analysis_real <inputFitsFile> <lambda> <J_0> <multiresFlag>
>> ./bin/s2let_axisym_mw_synthesis_real <outputRoot> <lambda> <J_0> <bandLimit>
\end{lstlisting}\end{minipage}\end{figure}

\begin{figure}[h!]\begin{minipage}{0.95\linewidth}\begin{lstlisting}[label=codeBash2, caption=Performing the forward (analysis) and inverse (synthesis) wavelet transform of a real signal ({\tt HEALPix} sampling) from the command line.]
 >> ./bin/s2let_axisym_hpx_analysis_real <inputFitsFile> <lambda> <J_0> <bandLimit>
 >> ./bin/s2let_axisym_hpx_synthesis_real <outputRoot> <lambda> <J_0> <bandLimit>
\end{lstlisting}\end{minipage}\end{figure}

\subsection{Wavelet transform in {\tt Matlab} and {\tt IDL}}

Examples \ref{codeMatlab} and \ref{codeIDL} read real signals on the sphere from {\tt FITS} files, calculate the wavelet coefficients and plot them using a Mollweide projection. The first case is a {\tt Matlab} example where the input map is a simulation of the cosmic microwave background in the {\tt HEALPix} sampling. The second case is a {\tt IDL} example where the input map is a topography map of the Earth in MW sampling. {\tt S2LET} ships with versions of these two examples in {\tt C}, {\tt Matlab} and {\tt IDL}. 

\begin{figure}[]\begin{minipage}{0.95\linewidth}\begin{lstlisting}[language=Matlab, label=codeMatlab, caption=Performing the wavelet transform of a real signal ({\tt HEALPix} sampling) using the {\tt Matlab} interface.]
% Example: Wavelet transform in Matlab
lambda = 3; J0 = 2; L = 192; 
Jmax = s2let_jmax(L, lambda);

% Read a real HEALPix map from a FITS file
inputfile = 'data/somecmbsimu_hpx_128.fits';
[f, nside] = s2let_hpx_read_real_map(inputfile);

% Perform the wavelet transform
[f_wav, f_scal] = s2let_axisym_hpx_analysis (f,'B',lambda,'L',L,'J_min',J0);

% Plot the map and the wavelet coefficients
figure; ns = ceil(sqrt(2+Jmax-J0+1));
subplot(ns, ns, 1);
s2let_hpx_plot_mollweide(f);
title('Initial band-limited data')
subplot(ns, ns, 2);
s2let_hpx_plot_mollweide(f_scal);
title('Scaling fct')
for j = J0:Jmax
   subplot(ns, ns, j-J0+3);
   s2let_hpx_plot_mollweide(f_wav{j-J0+1});
   title(['Wavelet scale : ',int2str(j)-J0+1])
end 
\end{lstlisting}\end{minipage}\end{figure}

\begin{figure}[]\begin{minipage}{0.95\linewidth}\begin{lstlisting}[language=IDL, label=codeIDL, caption=Performing the wavelet transform of a real signal (MW sampling) using the {\tt IDL} interface.]
; Example: Wavelet transform in IDL
lambda = 3
J0 = 2

; Read a real MW map from a FITS file
file = 'data/earth_tomo_mw_128.fits'
f = s2let_mw_read_real_map(file)
L = s2let_get_mw_bandlimit(f)
Jmax = s2let_j_max(L, lambda)

; Perform the wavelet transform
f_wav = s2let_axisym_mw_wav_analysis_real (f, lambda, J0)
f_rec = s2let_axisym_mw_wav_synthesis_real (f_wav)

; Plot the map and the wavelet coefficients
ns = ceil(sqrt(3+Jmax-J0)) 
!P.MULTI=[0,ns,ns]
s2let_mw_plot_mollweide, f_rec, title='Band-limited map'
s2let_mw_plot_mollweide, f_wav.scal, title='Scaling map'
for j=0, Jmax-J0 do begin
   s2let_mw_plot_mollweide, f_wav.(j), title='Wavelet map '+strtrim(j+1,2)+' on '+strtrim(Jmax-J0+1,2)
endfor
!P.MULTI=0
\end{lstlisting}\end{minipage}\end{figure}

\subsection{Wavelet denoising in {\tt C}}

Example \ref{codeC} illustrates the use of the wavelet transform to denoise a signal on the sphere. The input noisy map is a band-limited topography map of the Earth in MW sampling at resolution $L=128$. It is read from a {\tt FITS} file, decomposed into wavelet coefficients (for given parameters $\lambda$ and $J_0)$ which are then denoised by thresholding. The denoised signal is reconstructed from the denoised wavelet coefficients and written to a {\tt FITS} file.

\begin{figure}[]\centering
\subfigure[Band-limited signal]{\includegraphics[trim = 1.8cm 23.5cm 0.7cm 1.4cm, clip, width=8.cm]{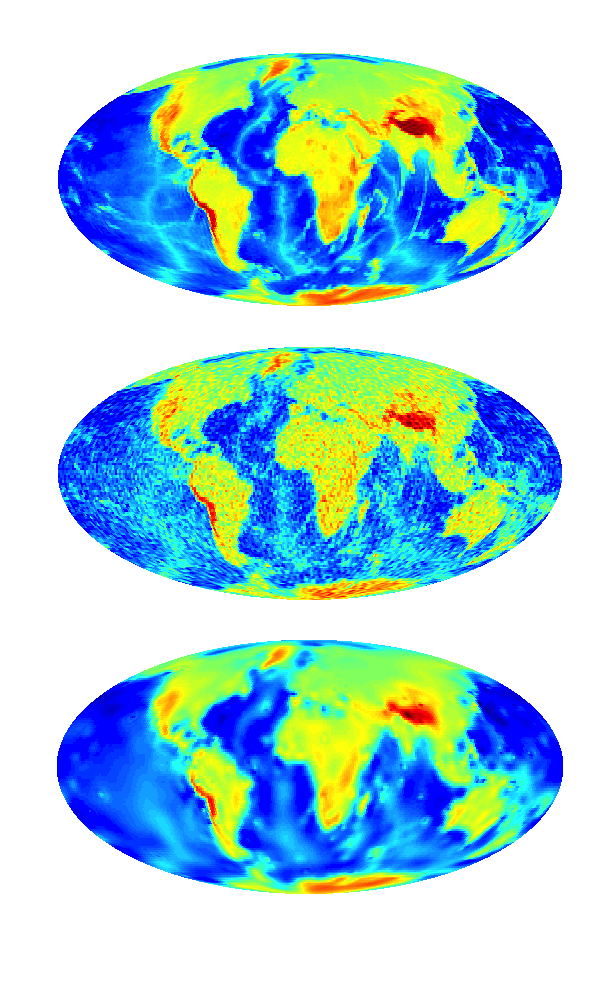}}
\subfigure[Noisy signal  with \mbox{$\textrm{SNR}(y)=11.8$dB}]{\includegraphics[trim = 1.8cm 13.1cm 0.7cm 11.9cm, clip, width=8.cm]{pics/denoising.png}}
\subfigure[Denoised signal with \mbox{$\textrm{SNR}(d)=14.66$dB}]{\includegraphics[trim = 1.8cm 2.7cm 0.7cm 22.2cm, clip, width=8.cm]{pics/denoising.png}}
\caption{Wavelet denoising by hard-thresholding, using parameters $\lambda=2$ and $J_0=0$ and scale-discretised generating functions.{When using needlets and B-spline wavelets, the denoised signals have \mbox{$\textrm{SNR}(d)=14.68$dB} and \mbox{$14.46$dB} respectively. This example is included in {\tt S2LET} as a documented demo program.}}
\label{fig:denoising}
\end{figure}

In this example we consider a noisy signal $y=s+n \in L^2(S^2)$, where the signal of interest $s\in L^2(S^2)$ is contaminated with noise $n\in L^2(S^2)$. We consider zero-mean white Gaussian noise on the sphere, where the variance of the harmonic coefficients of the noise is specified by
\begin{equation}
	 \mathbb{E}\left( |{n}_{\ell m}|^2 \right) \ = \ \sigma^2, \quad \forall \ell, m. \label{noisevar}
\end{equation}
A simple way to evaluate the fidelity of the observed signal $y$ is through the signal-to-noise ratio (SNR), define on the sphere by
\begin{equation}
	\textrm{SNR}(y) \equiv 10 \log_{10} \frac{ \| s \|_2^2 }{ \| y - s \|_2^2 },
\end{equation}
where the signal energy is defined by
\begin{equation}
	\| y \|_2^2 \ \equiv \  \langle y|y \rangle= \ \int_{S^2} d\Omega(\omega)  | y(\omega) |^2  \ = \ \sum_{\ell m} | y_{\ell m} |^2.
\end{equation}
We seek a denoised version of $y$, denoted by $d \in L^2(S^2)$, with large ${\rm SNR}(d)$ so that $d$ isolates the informative signal $s$.  When taking the wavelet transform of the noisy signal $y$, one expects the energy of the informative part to be concentrated in a small number of wavelet coefficients, whereas the noise energy should be spread over various wavelet scales.  {In this particular toy example, the signal has significant power on large scales, as shown in Figure~\ref{fig:earth}, which are well described in the wavelet basis and less affected by the random white noise.} Since the transform is linear, the wavelet coefficients of the $j$-th scale are simply given by the sum of the individual contributions:
\begin{equation}
	Y^{j}(\omega) = S^{j}(\omega) +  N^{j}(\omega),
\end{equation}
where capital letters denote the wavelet coefficients, i.e. \mbox{$Y^{{j}} \equiv y \star \Psi^{j}$}, $S^{{j}} \equiv s \star \Psi^{j}$ and $N^{{j}} \equiv n \star \Psi^{j}$.
For the zero-mean white Gaussian noise defined by Eqn. \ref{noisevar}, the noise in wavelet space is also zero-mean and Gaussian, with variance
\begin{equation}
	\mathbb{E}\left( |{N}^{j }(\omega)|^2 \right) = \sigma^2 \sum_{\ell} |  {\Psi}^{j}_{\ell 0} |^2    \ \ \label{noisemodel} \equiv  \left( \sigma^{j} \right)^2. \nonumber
\end{equation}
Denoising is performed by hard-thresholding the wavelet coefficients $Y^{j}$, where the threshold is taken as \mbox{$T^j = 3\sigma^{j}$}.  The denoised wavelet coefficients \mbox{$D^{{j}} \equiv d \star \Psi^{j}$} are thus given by
\begin{equation}
	D^{j }(\omega) = \left\{\begin{array}{ll}
		0 ,   & \textrm{if } Y^{j }(\omega)< T^j(\omega) \\
		Y^{j }(\omega),    &\textrm{otherwise}
		\end{array}\right. . \label{threshold}
\end{equation}
The denoised signal $d\in L^2(S^2)$ is reconstructed from its wavelet coefficients $D^{j }$ and the scaling coefficients of $y$, which are not thresholded.  The denoising procedure outlined above is particularly simple and more sophisticated denoising strategies can be developed; we adopt this simple denoising strategy merely to illustrate the use of the {\tt S2LET} code.  In this example we perform the wavelet transform with parameters $\lambda=2$ and $J_0=0$.  For a noisy signal $y$ with $\textrm{SNR}(y)=11.78$dB, the scale-discretised wavelet denoising recovers a denoised signal $d$ with \mbox{$\textrm{SNR}(d)=14.66$dB}.  The initial, noisy and denoised maps are shown in Figure~\ref{fig:denoising}. {When switching to needlets and B-spline wavelets while keeping $\lambda$ and $J_0$ unchanged, the denoised signals have \mbox{$\textrm{SNR}(d)=14.68$dB} and \mbox{$14.46$dB} respectively. }

\begin{figure}[]\begin{minipage}{0.95\linewidth}\begin{lstlisting}[label=codeC, caption=Denoising a real signal (MW sampling) in {\tt C} through hard-thresholding of the wavelet coefficients.]
// Example: Wavelet denoising in C
int lambda = 2, J0 = 0;

// Read a real MW map from a FITS file
char inputfile[100] = "..."
double *f;
int L = s2let_fits_read_mw_bandlimit(file);
s2let_mw_allocate_real(&f, L); 
s2let_fits_read_mw_map(f, file, L); 

// Perform multiresolution wavelet analysis
double *f_wav, *f_scal;
s2let_axisym_mw_allocate_f_wav_multires_real (&f_wav, &f_scal, lambda, L, J0);
s2let_axisym_mw_wav_analysis_multires_real (f_wav, f_scal, g, lambda, L, J0);

// Threshold the wavelets with a noise model
s2let_axisym_wav_hardthreshold_multires_real (f_wav, threshold, lambda, L, J0);

// Reconstruct the denoised signal
double *f_denoised;
s2let_mw_allocate_real(&f_denoised, L);
s2let_axisym_mw_wav_synthesis_multires_real (f_denoised, f_wav, f_scal, lambda,L,J0);

// Write the denoised signal
char outputfile[100] = "..."
s2let_fits_write_mw_map(outfile,f_denoised,L); 
\end{lstlisting}\end{minipage}\end{figure}

\section{Summary}

In the era of precision astrophysics and cosmology, large and complex data-sets on the sphere must be analysed at high precision in order to confront accurate theoretical predictions. Scale-discretised wavelets are a powerful analysis technique where spatially localised, scale-dependent signal features of interest can be extracted and analysed. Combined with a sampling theorem, this framework leads to an exact multiresolution wavelet analysis, where signals on the sphere can be reconstructed from their scaling and wavelet coefficients exactly.

We have described {\tt S2LET}, a fast and robust implementation of the scale-discretised wavelet transform. Although the first public release of {\tt S2LET} is restricted to axisymmetric wavelets, the generalisation to directional, steerable wavelets will be made available in a future release. The core numerical routines of {\tt S2LET} are written in {\tt C} and have interfaces in {\tt Matlab}, {\tt IDL} and {\tt Java}. Both MW and {\tt HEALPix} pixelisation schemes are supported. In this article we have presented a number of examples to illustrate the ease of use of {\tt S2LET} for performing wavelet transform of real signals stored as {\tt FITS} files and to plot scaling and wavelet coefficients on Mollweide projections of the sphere. We have also detailed a denoising example where denoising is performed through simple hard-thresholding in wavelet space.  Although only a simple denoising strategy was performed to illustrate the use of the {\tt S2LET} code, it nevertheless performed very well, highlighting the effectiveness of the scale-discretised wavelet transform on the sphere.

\bibliographystyle{aa}
\bibliography{biblio}

\begin{acknowledgements}
BL is supported by the Perren Fund and the Impact Fund. JDM is supported in part by a Newton International Fellowship from the Royal Society and the British Academy. YW is supported by the Center for Biomedical Imaging (CIBM) of the Geneva and Lausanne Universities, EPFL and the Leenaards and Louis-Jeantet foundations.
\end{acknowledgements}

\end{document}